# Relativistic Artificial Molecules Realized by Two Coupled Graphene Quantum Dots

Zhong-Qiu Fu[1,§], Yue-Ting Pan[1,§], Jiao-Jiao Zhou[2,§], Dong-Lin Ma[1,3], Yu Zhang[1], Jia-Bin Qiao[1], Haiwen Liu[1,*], Hua Jiang[2,*], and Lin He[1,*]

[1] Center for Advanced Quantum Studies, Department of Physics, Beijing Normal University, Beijing, 100875, People's Republic of China

[2] School of Physical Science and Technology and Institute for Advanced Study, Soochow University, Suzhou, 215006, People's Republic of China

[3] Department of Physics, Capital Normal University, Beijing, 100048, People's Republic of China

[§]These authors contributed equally to this work.

[*]Correspondence and requests for materials should be addressed to H.L (haiwen.liu@bnu.edu.cn), H.J (e-mail: jianghuaphy@suda.edu.cn), and L.H. (e-mail: helin@bnu.edu.cn).

**Coupled quantum dots (QDs), usually referred to as artificial molecules, are important not only in exploring fundamental physics of coupled quantum objects, but also in realizing advanced QD devices. However, previous studies have been limited to artificial molecules with nonrelativistic fermions. Here, we show that relativistic artificial molecules can be realized when two circular graphene QDs are coupled to each other. Using scanning tunneling microscopy (STM) and spectroscopy (STS), we observe the formation of bonding and antibonding states of the relativistic artificial molecule and directly visualize these states of the two coupled graphene QDs. The formation of the relativistic molecular states strongly alters distributions of massless Dirac fermions confined in the graphene QDs. Because of the relativistic nature of the molecular states, our experiment demonstrates that the degeneracy of different angular-momentum states in the relativistic artificial molecule can be further lifted by external magnetic fields. Then, both the bonding and antibonding states are split into two peaks.**

Quantum dots (QDs), which are usually called artificial atoms, can confine electrons to quantized electronic states with discrete energies. Therefore, two coupled QDs are often referred to as an artificial molecule [1-7]. The study of the coherent superposition and entanglement of localized quantum-confined states in these artificial molecules has attracted much attention over the years [1-25], not least because they are important to understand the fundamental physics of coupled quantum objects, but also because they have potential applications in future quantum information processing. Although enormous progress has been made in the past few decades, only artificial molecules with nonrelativistic fermions have been studied up to now. In this Letter, we demonstrate experimentally that a relativistic artificial molecule can be realized when two graphene QDs are coupled to each other. The bonding and antibonding states of the relativistic artificial molecule are directly imaged by using scanning tunneling microscope (STM). Because of the relativistic nature of the molecular states, we show that magnetic fields can lift the degeneracy of different angular-momentum states in the relativistic artificial molecule.

Figure 1 shows schematic images of three "generations" of molecules. The first generation is a real molecule formed by two real atoms. Figure 1(a) schematically shows a hydrogen molecule, where the molecular states, *i.e.*, the bonding and antibonding levels, are formed because of the coupling of two hydrogen atoms. The second generation is quantum-dot molecule with tunable coupling strength between the two artificial atoms (Figure 1(b)). Both the first and second generations are nonrelativistic molecules. The third generation is relativistic artificial molecule realized by two coupled graphene QDs (Figure 1(c)), as first proposed in this work. In a graphene QD defined by a circular *p-n* junction, massless Dirac fermions, *i.e.*, the so-called relativistic fermions, of graphene can be temporarily trapped [26-29]. Because of the unusual anisotropic transmission across the *p-n* junction (the Klein tunneling in graphene) [30], the massless Dirac fermions are temporarily confined into quasibound states in the graphene QDs via whispering-gallery-mode (WGM) confinement [26]. We will demonstrate subsequently that the coupling of the quasibound states in two adjacent graphene QDs, as schematically shown in Fig. 1(c), could lead to the formation of relativistic molecular states.

To obtain coupled graphene QDs, a new method is developed to generate circular *p-n* junctions in a continuous graphene sheet. In our experiment, graphene monolayer was synthesized at high temperature by low pressure chemical vapor deposition (LPCVD) on a S-rich copper foil [31]. Then the sample is slowly cooled down to room temperature. The S atoms segregate from the Cu foil onto the metal surface and form ordered S superlattice during the cooling process [31]. Our experiment in this work demonstrates that abundant S atoms will further assemble into nanoscale dots on top of the ordered S superlattice formed advanced. The nanoscale S dots intercalate the interface between the graphene sheet and the Cu foil (see Fig. S1 for a representative STM image of a region showing several nanoscale S dots [32]). The thickness of the S dots is measured to be about 0.2 nm (see Fig. S2 [32]), consisting with that of monolayer S atoms on Cu surface [33,34]. The atomic-layer difference of graphene-Cu separations between inside and outside of the nanoscale S dots introduce sharp electronic junctions in the continuous graphene monolayer. To verify this, we carry out field-emission resonances (FER) measurements of graphene on and off the nanoscale S dot region (see Fig. S3 [32]). According to the energy shifts in the first FER peak, the difference in the local work function of graphene on and off the S dot region is measured to be about 300 meV. Our high-field scanning tunneling spectroscopy (STS) measurements obtain Landau quantization of the graphene monolayer off the S dot region, indicating that the ordered S superlattice effectively decouples the interaction between the graphene and the Cu foil. According to the measured Landau levels, the Dirac point $E_D$ of graphene off the QDs is measured as about -90 meV (see Fig. S4 for the details [32]). Therefore, the Dirac point of graphene in the QDs should be at about 210 meV. The formation of *p-n* junctions in graphene along the edges of the S dots is further confirmed by carrying out STS measurements. A series of equally spaced quasibound states arising from the WGM-type confinement are observed in the tunneling spectra recorded in the graphene QD (see Fig. S5 for the details [32]), indicating the quantum confinement of the massless Dirac fermions by the circular *p-n* junction of the graphene QD.

Taking advantage of a great number of graphene QDs formed in the sample, it is facile to find two adjacent graphene QDs to study the relativistic artificial molecular states. Figure 2(a) shows a representative STM image of two coupled graphene QDs. The schematic side view of the two coupled QDs along with the corresponding diagram of the two *p-n-p* junctions is shown in Fig. 2(b). The two adjacent graphene QDs almost have identical size and structure, which ensure that the energy spectra of the quasibound states in the two QDs are similar. Here we should point out that the electrically decoupling between the graphene and the Cu foil is very important for the coupling of two adjacent graphene QDs, *i.e.*, for the realization of the relativistic artificial molecular states. Our STS spectra recorded in the two graphene QDs, as shown in Fig. 2(c), demonstrate explicitly that the energies of the quasibound states in the two graphene QDs are almost the same. The average energy spacing of the quasibound states ~ 120 meV agrees well with that estimated according to $\Delta E \approx \hbar v_F/R$, where $\hbar$ is the reduced Planck's constant, $v_F = 1.0 \times 10^6$ m/s is the Fermi velocity of graphene monolayer, and $R = \sqrt{A/\pi}$ is the effective radius of the QDs with *A* the area of each QD measured in STM images. Such a result, which is similar as that of isolated graphene QDs [26-29], indicates the quantum confinement of the massless Dirac fermions in the two coupled graphene QDs via the WGM modes.

However, there is an obvious difference between the spectra recorded in the two coupled graphene QDs and that in an isolated graphene QD: the lowest quasibound state measured in the two coupled QDs is split into two peaks with an energy separation of about 30 meV, as shown in high-resolution spectra in Figs. 2(d) and 2(e). Such a result reminds us the formation of molecular states, *i.e.*, the bonding and antibonding states (σ and σ*), of the coupled QDs [1-7]. In an isolated graphene QD, the lowest quasibound state, which exhibits a maximum in the center of the QD, is not well confined in the QD because it is not a WGM mode [26]. For a fixed radial momentum, the quasibound states with higher angular momentum (higher energy) are much better confined due to the formation of high-finesse WGM resonances. In addition, the wavelength of the quasiparticles in graphene monolayer increases with decreasing the energy. Therefore, in the coupled graphene QDs, the coupling strength is strongest for the lowest quasibound state and is expected to decrease with increasing the energy. In our experiment, the coupling strength, as reflected by the splitting, is much stronger for the lowest quasibound state and decrease with increasing the energy of the quasibound state, as shown in Fig. 2(c). Such a phenomenon is a characteristic feature of the formation of the relativistic artificial molecule in the two graphene QDs. We can directly image the bonding and antibonding states of the coupled QDs by operating energy-fixed STS mapping, which reflects the local density of states (LDOS) in real space. Figure 2(f) and 2(g) show the LDOS spatial maps of the two split peaks, which exhibit quite different features obtained in an isolated graphene QD (see Fig. S5 [32]) and directly demonstrate the formation of a bonding and an antibonding molecular state in the two coupled graphene QDs.

To further confirm the formation of the relativistic molecular states, we carried out theoretical calculations of a coupled graphene QDs in a continuous graphene sheet based on the lattice Green's function [29,35]. For simplicity, we consider two circular graphene QDs with the radius of the QD as $R = 6$ nm, the distance between the two QDs as $d = 4$ nm, and the potential difference on and off the QDs as 270 meV, as schematically shown in Fig. 3(a). Figure 3(b) shows the theoretical LDOS curves of the two coupled graphene QDs at different positions, which capture well the main features of the experimental result in several aspects. First, the theoretical spectra exhibit a series of quasibound states with almost the same energy separations between the quasibound states as observed in our experiment. It indicates that these quasibound states are arising from the confinement of the "relativistic Dirac fermions" in graphene monolayer. Second, the lowest quasibound states split into two peaks, as marked by $\sigma$ and $\sigma^*$, because of the coupling between the two adjacent QDs. It is well known that two QDs will form artificial molecule when they are close and, then, their confined states will split into a bonding orbit with lower energy and an antibonding orbit with higher energy (Although the splitting energy depends on the distance of the two QDs, the main feature of the splitting is not affected, as shown in Fig. S6 [32]). Here, the split of the lowest quasibound states, which agrees with our experimental result, proves the formation of the artificial molecular states in the coupled graphene QDs. Third, the split of the quasibound states decreases with increasing angular momentum (or energy), which is also well consistent with our experimental result. We obtain such a result, both experimentally and theoretically, because that the quasibound states with higher angular momentum (higher energy) are much better confined in each QD due to the formation of high-finesse WGM resonances. The spatial distributions of the LDOS of the artificial molecular states in the coupled graphene QDs are also calculated, as shown in Fig. 3(c) and 3(d). Theoretically, the wavefunctions of the bonding and antibonding states, $\psi_\sigma$ and $\psi_{\sigma*}$, of the two coupled QDs can be approximated by the sum and difference of the wavefunctions of individual dots respectively, as shown in Fig. 3(e). Therefore, the spatial distributions of the LDOS for the bonding state and antibonding state are quite different. Obviously, the obtained theoretical result, as shown in Fig. 3(c) and 3(d), agrees quite well with that observed in our experiment, as shown in Fig. 2(f) and 2(g).

The relativistic molecular states in the coupled graphene QDs is further explored by carrying out high-magnetic-field STS measurements. Figure 4(a) shows tunneling spectra measured at the center of the left QD, as shown in Fig. 2(a), in different magnetic fields. It is interesting to note that the two molecular states gradually split into four peaks, as marked by $\sigma_+^*, \sigma_-^*, \sigma_+, \sigma_-$, with increasing the applied magnetic fields. By measuring the position of these four peaks as a function of magnetic fields, we find that the energy separation between the bonding and antibonding states almost does not change with the magnetic field, however, the splitting between $\sigma_+^*$ ($\sigma_+$) $and$ $\sigma_-^*$ ($\sigma_-$) increases linearly with magnetic field, as shown in Fig. 4(c). Here, we attribute the large splitting to the lifting of the degeneracy with opposite angular momentum quantum number $\pm m$ of the quasibound states in graphene QDs. The fourfold degeneracy of the first quasibound state also proves the formation of the artificial molecule because that the molecular states increase the number of electrons contained in the first quasibound state. A perpendicular magnetic field bends the trajectory of electrons and break the time reversal symmetry in graphene QDs, which consequently lifts the degeneracy of the $\pm m$ sublevels in the quasibound states [36-41]. Very recently, it was demonstrated explicitly that the magnetic fields will lead to a sudden jump in phase by $\pi$ of the quasiparticles as the sign of the incident angle is changed by a critical magnetic field. Such an effect results in a larger separation of the $\pm m$ sublevels than the separation in our experiment because we cannot reach the critical magnetic field (about tens Tesla in our system) [37-40]. In our experiment, the magnetic field induced split of $\pm m$ increases linearly with magnetic field. By fitting the linear relationship between the split and the magnetic fields ($\Delta E = g^*\mu_B B$), we obtain the effective $g$ factor $g^* \approx 40$ for the orbital magnetic moment, which is about 20 times larger than that arises from the Zeeman effect of spin magnetic moment (the $g$ factor for spin is 2).

To further understand the effect of magnetic fields on the relativistic molecular states, we calculate the influence of magnetic fields on the bonding and antibonding states of two coupled graphene QDs by Wentzel Kramers Brillouin (WKB) approximation. In the calculation, the coupling strength between the two QDs is assumed to be a constant because the split of bonding and antibonding states is almost unchanged in our experiment. The calculated dI/dV curves are shown in Fig. 4(b), and the $\pm m$ sublevels will split in magnetic field. Comparing the experimental and theoretical result, we can find that the left peak of bonding and antibonding states (marked by $\sigma^*+$ and $\sigma+$) are slightly higher than the right peak of bonding and antibonding states (marked by $\sigma^*-$ and $\sigma-$), which is caused by the anisotropic influence by magnetic field on $\pm m$ sublevels. At last, we plot the $\pm m$ energy splitting of bonding and antibonding states as the function of magnetic field in Fig. 4(c). Our experimental values are slightly larger than the theoretical calculation, which may be caused by the imperfect circle in our experiment.

In conclusion, we study the electronic properties of two coupled graphene QDs and demonstrate the realization of relativistic artificial molecular states in such a unique system. The coupling of the quasibound states of the two QDs leads to the formation of the bonding and antibonding states, which are directly imaged in our experiment. Our experiment further demonstrate that the magnetic fields can lift the angular-momentum degeneracy of molecular states and, consequently, both the bonding and antibonding states of the relativistic artificial molecule are split.

## Acknowledgements

This work was supported by the National Natural Science Foundation of China (Grant Nos. 11674029). L.H. also acknowledges support from the National Program for Support of Top-notch Young Professionals, support from "the Fundamental Research Funds for the Central Universities", and support from "Chang Jiang Scholars Program".## References

[1]  G. Schedelbeck, W. Wegscheider, M. Bichler, and G. Abstreiter, Coupled Quantum Dots Fabricated by Cleaved Edge Overgrowth: From Artificial Atoms to Molecules. Science **278**, 1792 (1997).

[2]  R. H. Blick, D. Pfannkuche, R. J. Haug, K. v. Klitzing, and K. Eberl, Formation of a Coherent Mode in a Double Quantum Dot. Phys. Rev. Lett. **80**, 4032 (1998).

[3]  B. Partoens and F. M. Peeters, Molecule-Type Phases and Hund's Rule in Vertically Coupled Quantum Dots. Phys. Rev. Lett. **84**, 4433 (2000).

[4]  D. Dixon, L. P. Kouwenhoven, P. L. McEuen, Y. Nagamune, J. Motohisa, and H. Sakaki, Influence of energy level alignment on tunneling between coupled quantum dots. Phys. Rev. B **53**, 12625 (1996).

[5]  F. R. Waugh, M. J. Berry, C. H. Crouch, C. Livermore, D. J. Mar, R. M. Westervelt, K. L. Campman, and A. C. Gossard, Measuring interactions between tunnel-coupled quantum dots. Phys. Rev. B **53**, 1413 (1996).

[6]  T. H. Oosterkamp, T. Fujisawa, W. G. van der Wiel, K. Ishibashi, R. V. Hijman, S. Tarucha, and L. P. Kouwenhoven, Microwave spectroscopy of a quantum-dot molecule. Nature **395**, 873 (1998).

[7]  A. W. Holleitner, R. H. Blick, A. K. Hüttel, K. Eberl, and J. P. Kotthaus, Probing and Controlling the Bonds of an Artificial Molecule. Science **297**, 70 (2002).

[8]  N. J. Craig, J. M. Taylor, E. A. Lester, C. M. Marcus, M. P. Hanson, and A. C. Gossard, Tunable Nonlocal Spin Control in a Coupled-Quantum Dot System. Science **304**, 565 (2004).

[9]  T. Unold, K. Mueller, C. Lienau, T. Elsaesser, and A. D. Wieck, Optical Control of Excitons in a Pair of Quantum Dots Coupled by the Dipole-Dipole Interaction. Phys. Rev. Lett. **94**, 137404 (2005).

[10]  A. P. Alivisatos, Semiconductor Clusters, Nanocrystals, and Quantum Dots. Science **271**, 933 (1996).

[11]  R. E. Bailey and S. Nie, Alloyed Semiconductor Quantum Dots: Tuning the Optical Properties without Changing the Particle Size. J. Am. Chem. Soc. **125**, 7100 (2003).

[12]  A. D. Yoffe, Semiconductor quantum dots and related systems: Electronic, optical, luminescence and related properties of low dimensional systems. Adv. Phys. **50**, 1 (2001).

# Figures

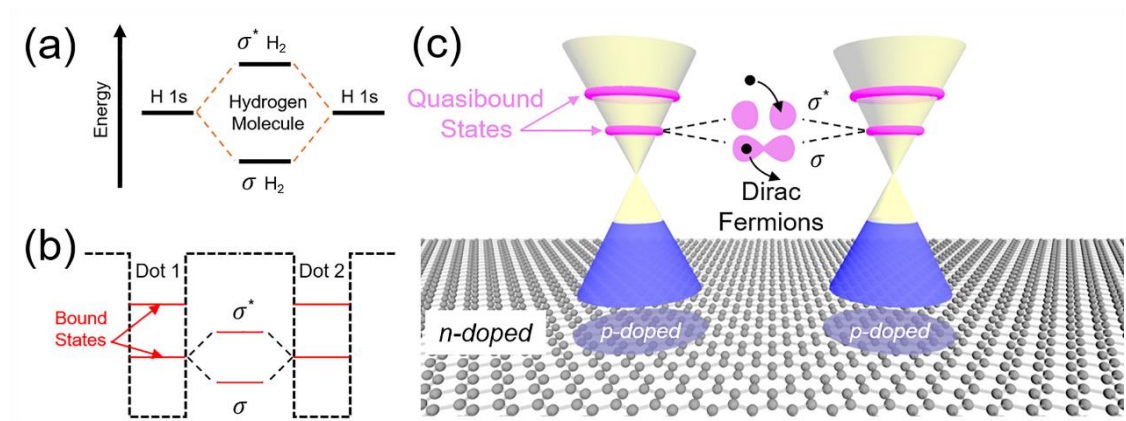

**Figure 1** Illustrations from hydrogen molecule to graphene quantum dots molecule. **(a)** The formation of hydrogen molecule from two hydrogen atoms. **(b)** The interaction between two classical quantum dots and the formation of quantum dots molecule. **(c)** The formation of quasibound molecular states in DGQD system.

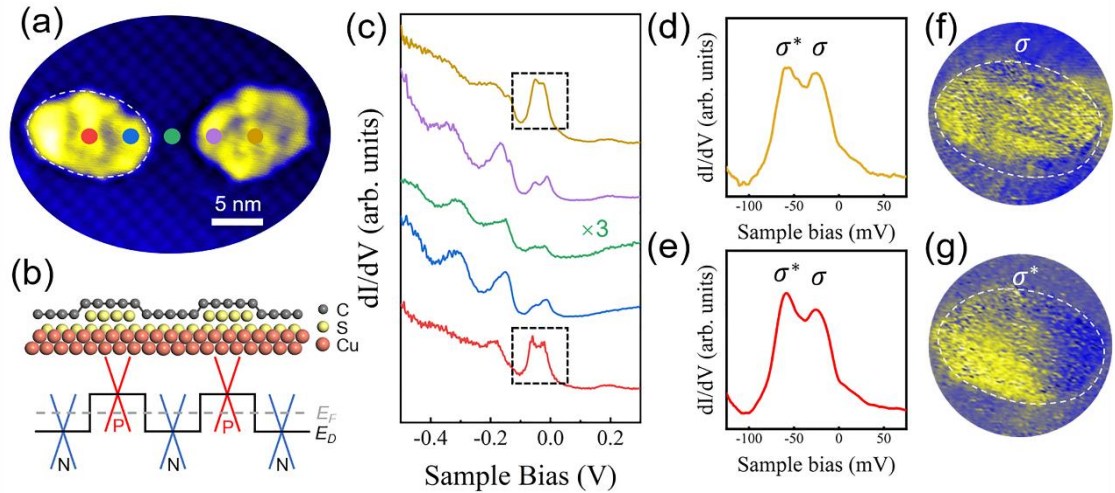

**Figure 2** The formation of quasibound molecular states and their corresponding LDOS. **(a)** Typical STM image of the double graphene quantum dots system. Scale bar: 5 nm. **(b)** Schematic diagram of the p-n-p junction. **c.** Tunneling spectra measured in different position marked in panel (a). **(d, e)** High-resolution tunneling spectra near the first quasibound state measured in the center of each quantum dot (dash squares in (**c**)). **(f, g)** STS maps recorded in the left quantum dot with the energy of $\sigma$ and $\sigma^*$ states.

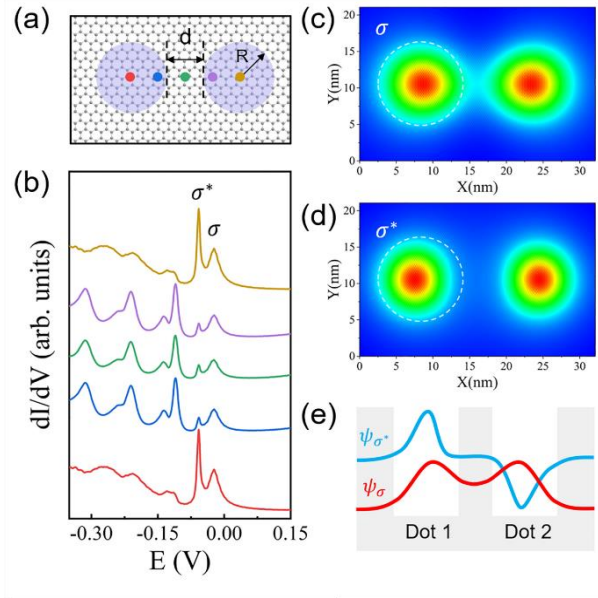

**Figure 3** Theoretical calculation of energy level and LDOS in double quantum dots system. **(a)** Schematic representation of the theory calculation model. **(b)** The calculated tunneling curves at different positions marked by the solid dots with different colors in the top panel of panel (a). **(c, d)** Theoretical spatial distribution of the LDOS with the energies σ and σ* states. **(e)** Schematic formation of wavefunction in our system.

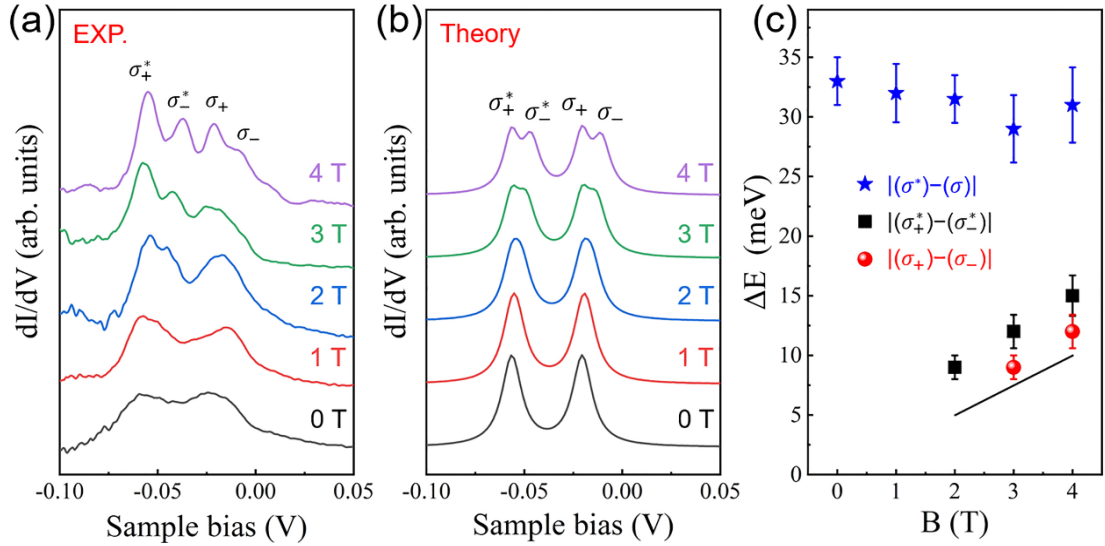

**Figure 4** The evolution of quasibound molecular states in magnetic field. **(a)** High-resolution STS spectra taken at the center of the left quantum dot under different magnetic fields. **(b)** The corresponding calculated tunneling curves under different magnetic fields. **(c)** The energy splitting of bonding and antibonding states (marked by blue star), $\pm m$ of bonding state (marked by red cycle) and $\pm m$ of antibonding state (marked by black square) as the function of magnetic field.